\documentclass[prb,aps,twocolumn, amssymb,amsmath,floatfix,superscriptaddress]{revtex4}
\usepackage{tabularx}
\usepackage{bm}
\usepackage{euscript}
\usepackage{epsfig,psfrag,subfigure}
\usepackage{graphicx}
\usepackage{color}
\usepackage{amsfonts}
\usepackage{exscale}
\usepackage{wrapfig}
\usepackage{extarrows} 
\usepackage{placeins}
\usepackage{verbatim}
\usepackage{amsmath}
\usepackage{enumitem}

\usepackage{hyperref}
\hypersetup{
    colorlinks=true,       
    linkcolor=red,          
    citecolor=blue,        
    filecolor=magenta,      
    urlcolor=cyan           
}

\usepackage{slashed} 
\numberwithin{equation}{section} 
\renewcommand{\theequation}{\arabic{section}.\arabic{equation}}

\newcommand{\bea}{\begin{eqnarray}}  
\newcommand{\eea}{\end{eqnarray}}
\newcommand{\ben}{\begin{enumerate}}
\newcommand{\een}{\end{enumerate}}
\newcommand{\be}{\begin{equation}}
\newcommand{\ee}{\end{equation}}

\renewcommand{\theequation}{\arabic{section}.\arabic{equation}}

\def\be{\begin{equation}}       \def\ee{\end{equation}}
\def\bea{\begin{eqnarray}}      \def\eea{\end{eqnarray}}
\def\ba{\begin{array} }
\def\ea{\end{array} }
\def\bnum{\begin{enumerate} }
\def\enum{\end{enumerate}}

\def\=>{\Rightarrow}
\def\>{\rightarrow}

\def\eye2{Fathbb{I}}

\def\Q{\vec{Q}}

\def\K{\vec{K}}

\def\r{\vec{r}}
\def\d0{\Delta_{0}}

\begin{document}

\title{Vestigial nematicity from spin and/or charge order in the cuprates} 

\author{Laimei Nie}
\affiliation{Department of Physics, Stanford University, Stanford, California 93105, USA}
\author{Akash  V. Maharaj}
\affiliation{Department of Physics, Stanford University, Stanford, California 93105, USA}
\author{ Eduardo Fradkin}
\affiliation{Department of Physics and Institute for Condensed Matter Theory,  University of Illinois, 1110 West Green Street, Urbana, Illinois 61801-3080, USA}
\author{Steven A. Kivelson}
\affiliation{Department of Physics, Stanford University, Stanford, California 93105, USA}
\date{\today}

\begin{abstract}
Nematic order has manifested itself in a variety of materials in the cuprate family.  We propose an effective field theory of a layered system with incommensurate, intertwined spin- and charge-density wave (SDW and CDW) orders, each of which consists of two components related by $C_4$ rotations. Using a variational method (which is exact in a large $N$ limit), we study the development of nematicity from partially melting those density waves by either increasing temperature or adding quenched disorder.  As temperature decreases we first find a transition to a nematic phase, but depending on the range of parameters (e.g. doping concentration) the strongest fluctuations associated with this phase reflect either proximate SDW or CDW order.  We also discuss the changes in parameters that can account for the differences in the SDW-CDW interplay between the (214) family  and the other hole-doped cuprates.
\end{abstract}

\maketitle

\section{Introduction}

One of the major developments of the last decade of research into the properties of the high temperature superconducting cuprates is the discovery that charge and spin density wave (CDW and SDW) ordering phenomena are ubiquitous.
\cite{tranquada1994simultaneous, tranquada1995evidence, kivelson-rmp2003, chang2012direct, ghiringhelli2012Long, rossat1991neutron, tranquada1992neutron, DaiYBCOspinGap-2001, howald2003periodic, hoffman2002four, parker2010fluctuating, comin2014charge, tabis2014charge, hayden2004structure, tranquada2004quantum} 
Moreover, it has been shown  that they strongly influence the electronic structure of these materials and 
have a significant (although possibly complex) relationship with the superconducting order.\cite{wakimoto-2003, LiTranquadaPRL, hucker2011stripe, liang2006evaluation, cyr2015suppression, chang2012direct, wu2013emergence,  Hamidian-2015, sonier-2007, taillefer-review-2010} This complex ``intertwining'' of multiple density-wave and superconducting orders\cite{fradkin2015intertwined} has long been  documented  in one family - the  214 family e.g. La$_{2-x}$Sr$_x$CuO$_4$ (LSCO)   - of hole-doped cuprates. 
There is also increasingly compelling evidence that various forms of order that break point-group symmetries but not translational symmetry may occur in an even broader range of parameters ({\it i.e.} temperature, doping, crystal structures,
etc.).\cite{ando-2002, achkar2016nematicity, lawler2010intra, fujita2014direct, hinkov2008electronic, daou-2010, louis2nematics, chang2011nernst, wu2015incipient, xia2008polar, fauque2006magnetic, li2008unusual, caprara2015signatures, zhao2016global, pelc2016unconventional, lubashevsky2014optical, bozovicPrivate}
Of these,  Ising nematic order, {\it i.e.} the spontaneous breaking of either a $C_4$ symmetry to $C_2$ or the breaking of a mirror symmetry which occurs as a result of a partial melting of a unidirectional CDW or SDW phase, is the most directly related to these same developments.\cite{kivelson-rmp2003}

These developments bring with them a host of associated new questions.  Among other things, one would like to understand the nature of the interplay between the CDW and SDW orders.  In YBa$_2$Cu$_3$O$_{6+\delta}$
(YBCO), a member of the 123 family of hole doped cuprates, they are apparently mutually exclusive\cite{keimerZn, Blanco-Canosa-Xray};  significant SDW correlations are observed at relatively low doped hole concentrations, $\delta < \delta_c \sim 8$\%,  while significant CDW correlations are only observed at higher doping, $\delta_c < \delta 
$, where there is a significant spin-gap and correspondingly little in the way of long-distance spin correlations.  In contrast, in the 214 materials SDW and CDW, correlations seem to grow cooperatively, their ordering vectors are apparently locked to each other, and they exist as readily detectable fluctuating order over a very broad range of doping.\cite{kivelson-rmp2003, tranquada-2007}     
In addition, in YBCO, \cite{ando-2002, achkar2016nematicity, lawler2010intra, fujita2014direct, hinkov2008electronic, daou-2010, louis2nematics, chang2011nernst, wu2015incipient}  there is a strongly nematic region of the phase diagram in a range of temperatures in which neither SDW nor CDW order is well developed, but it has been suggested that there may be two distinct nematic phases - one associated with ``vestigial'' SDW and the other with vestigial CDW order.  
Furthermore, there is  evidence of nematicity in 214 materials in the fluctuating stripes regime. \cite{achkar2016nematicity,kivelson-rmp2003} Given that one never sees true long-range CDW or SDW order, it is also clearly important to understand the effect of quenched randomness - ``disorder'' - on all these properties.

In the present paper we consider the properties of the simplest Landau-Ginzburg-Wilson classical effective field theory of 
thermal fluctuations of the SDW and CDW order parameters in a layered (quasi 2D) system with tetragonal symmetry.  We include effects of disorder as a Gaussian random field coupled to the CDW order parameter.  To obtain controlled solutions including the effects of order parameter fluctuations, we solve the problem using the Feynman (``self-consistent gaussian'') variational approach.\cite{feynman}  An alternative way to view this is to consider the generalized version of this model in which both the CDW  (which is an $O(2)$ field) and the SDW  (which is an $O(2)\times O(3)$ field) are generalized to $O(N)$ fields, and the problem can then be solved exactly in the $N\to \infty$ limit.  It has previously been shown by us\cite{niesachdev} and others\cite{hayward2014angular} that such an approach  captures much of the physics of the actual physical problem.

Our principal results concerning the nature of the interplay between CDW, SDW, and nematic order are summarized in the calculated phase diagrams in the  figures below. Of particular note:  
1)  We find that there is a {\it single} nematic phase, spanning two distinct regimes separated by a crossover: in one, the nematicity can be viewed as vestigial SDW order (with corresponding strong local SDW correlations) whereas in the other it is associated with short-range CDW order.\cite{nie2014quenched}  
2)  An interesting feature associated with the most natural choices of parameters is that the dominant unidirectional SDW correlations are perpendicular to the dominant unidirectional CDW correlations. This is strikingly reminiscent of the case in YBCO, where the nematic axis is pinned by the weak symmetry breaking field imposed by the crystalline orthorhombicity, and where it is found that the preferred direction of the SDW ordering vector is perpendicular to the chain direction,\cite{keimer2010NJP} while based on high field experiments\cite{chang2016magnetic, jang2016ideal}, the preferred CDW ordering vector is parallel to the chains.  
3)  There is a special cubic term in the effective field theory that operates only when the CDW and SDW are mutually commensurate.  Our primary focus is on the far from commensurate case, where this term is negligible, but, as we will argue, it is likely the key to understanding the differences between the La based 214 materials and YBCO. 

This paper is organized as follows:  In Sec. \ref{model} we define the model and discuss the meaning of the various symmetry allowed terms. In Sec. \ref{variational}, we introduce the variational approach, and in Sec. \ref{results} we apply it to obtain the phase diagram of the stated model both with and without disorder.  In Sec. \ref{commensurability}, we  discuss the effects of the special cubic term,\cite{zachar1998landau} Eq. \eqref{cubic}, which is important only when the CDW and SDW are mutually commensurate, {\it i.e.} when $2\K\equiv \Q$, where $\K$ and $\Q$ are, respectively, the SDW and CDW ordering vectors, 
 and ``$\equiv$" means equal modulo a reciprocal lattice vector.   Finally, in Sec. \ref{discussion}, we discuss the relevance of our results to interpreting experiments in the cuprates and in Sec. \ref{remarks} we discuss some broader issues of perspective.


\section{The model}
\label{model}
We consider a quasi-2D (layered) lattice with tetragonal symmetry. The spin and charge densities at position $\r$ can be expressed as
\be
\rho(\r) = \overline \rho + \Big[ \rho_{\Q}(\r) e^{i \Q \cdot \r} + \rho_{\Q^\prime}(\r) e^{i \Q^\prime \cdot \r}  + \mbox{c.c.} \Big] + \cdots,
\ee
\be
{\bf S}(\r) = \Big[ {\bf S}_{\K}(\r) e^{i \K \cdot \r} + {\bf S}_{\K^\prime}(\r) e^{i \K^\prime \cdot \r} + \mbox{c.c.} \Big] + \cdots,
\ee
where ${\bf S}_{\K}$ and ${\bf S}_{\K^\prime}$ ($\rho_{\Q}$ and $\rho_{\Q^\prime}$) are slowly varying complex vector (scalar) fields corresponding to incommensurate SDW (CDW) order parameters. Here $\K$ and $\K^\prime$ ($\Q$ and $\Q^\prime$)  are wavevectors (assumed incommensurate) within the $xy$ plane, have the same magnitude and are related by $C_4$ rotations.  Note that $\rho_{\Q}(\r)=\rho^*_{-\Q}(\r)$, $\rho_{\Q^\prime}(\r)=\rho^*_{-\Q^\prime}(\r)$, ${\bf S}_{\K}(\r)={\bf S}^*_{-\K}(\r)$, and ${\bf S}_{\K^\prime}(\r)={\bf S}^*_{-\K^\prime}(\r)$ .

Keeping all terms to fourth order in the field amplitudes and second order in spatial derivatives that are consistent with translational symmetry and spin-rotational symmetry, the classical Ginzburg-Landau-Wilson effective field theory for this problem is
\begin{widetext}
\bea
&&H =  \int d\r \bigg\{ 
 {\cal H}_S+{\cal H}_\rho+{\cal H}_{3d}+{\cal H}_{S-\rho}  +{\cal H}_{com}+{\cal H}_{dis}\bigg\},\nonumber \\
&&{\cal H}_S=  
\frac{\kappa_{s\|}}{2} \Big[|\partial_x{\bf S}_{\K}|^2  +|\partial_y{\bf S}_{\K^\prime}|^2 \Big] + 
 \frac{\kappa_{s\perp}}{2} \Big[|\partial_y{\bf S}_{\K}|^2   +|\partial_x{\bf S}_{\K^\prime}|^2 \Big]    +\frac {\alpha_s} 2 \Big[ |{\bf S}_{\K}|^2  +|{\bf S}_{\K^\prime}|^2 \Big] 
+ \frac{u_s}{4} \Big[ |{\bf S}_{\K}|^2   +|{\bf S}_{\K^\prime}|^2 \Big]^2 
    \nonumber\\
&&\ \ \ \ \ \ \ \ \ \ \ \ \ \ \ \  + \frac{\gamma_s}{2}  |{\bf S}_{\K}|^2 |{\bf S}_{\K^\prime}|^2  + \frac{\tilde u_s}{4} \Big| {\bf S}_{\K}\times  {\bf S}^*_{\K}  +{\bf S}_{\K^\prime}\times  {\bf S}^*_{\K^\prime}  \Big|^2
+  \frac{\tilde\gamma_s}{2} \big[{\bf S}_{\K}\times  {\bf S}^*_{\K}\big]  \cdot\big[{\bf S}_{\K^\prime}\times  {\bf S}^*_{\K^\prime}\big]     \nonumber\\
&& H_\rho=  \frac{\kappa_{\rho\|}}{2} \Big[|\partial_x\rho_{\Q}|^2  +|\partial_y\rho_{\Q^\prime}|^2 \Big]+ 
 \frac{\kappa_{\rho\perp}}{2} \Big[|\partial_y\rho_{\Q}|^2   +|\partial_x\rho_{\Q^\prime}|^2 \Big] 
+\frac {\alpha_\rho} 2 \Big[ |\rho_{\Q}|^2 +|\rho_{\Q^\prime}|^2 \Big] + \frac{u_\rho}{4} \Big[ |\rho_{\Q}|^2 +|\rho_{\Q^\prime}|^2 \Big]^2  \nonumber \\
&&\ \ \ \ \ \ \ \ \ \ \ \ \ \ \ \ + \frac{\gamma_\rho}{2} |\rho_{\Q}|^2 |\rho_{\Q^\prime}|^2   \nonumber\\
&&{\cal H}_{3d}= -J_{sz} \Big[ {\bf S}_{\K}(n) \cdot  {\bf S}^*_{\K}(n+1) + {\bf S}_{\K^\prime}(n) \cdot  {\bf S}^*_{\K^\prime}(n+1) +c.c. \Big] - J_{\rho z} \Big[  \rho_{\Q}(n) 
  \rho^*_{\Q}(n+1) + \rho_{\Q^\prime}(n) 
 \rho^*_{\Q^\prime}(n+1)+ c.c.   \Big] \nonumber\\
 &&{\cal H}_{S-\rho}= \frac{v}{2} \Big[|\rho_{\Q}|^2|{\bf S}_{\K}|^2+|\rho_{\Q^\prime}|^2|{\bf S}_{\K^\prime}|^2\Big] + \frac {v'}{2} \Big[|\rho_{\Q}|^2|{\bf S}_{\K'}|^2+|\rho_{\Q^\prime}|^2|{\bf S}_{\K}|^2\Big] ,
\label{Hamiltonian}
\eea
\end{widetext}
where $\int d\vec r \equiv \sum\limits_n \int d x d y$ ($n$ labels the $z$-direction layers), and we have adopted the short-hand notation ${\bf S}_{\K} \equiv {\bf S}_{\K}(\vec r) = {\bf S}_{\K}(x,y,n)$ and ${\bf S}_{\K}(n) \equiv {\bf S}_{\K}(x,y,n)$ etc..   ${\cal H}_{com}$ and ${\cal H}_{dis}$ represent, respectively, a possible commensurate locking term between the CDW and SDW, and  the effect of disorder, both of which we discuss below.\cite{su2}


\subsection{Choice of less crucial parameters}
\label{parameters}
 There is a large number of parameters in this equation.  Many of them are not qualitatively important for the issues at hand, and so we will henceforth arbitrarily assume values chosen to simplify the ensuing  analysis.  We will take $\kappa_{s\parallel} = \kappa_{s\perp} =\kappa_{\rho\parallel} = \kappa_{\rho\perp} \equiv \kappa$, which is to say we will ignore anisotropy of the density wave elastic constants and the difference in the compressibilities of the CDW and SDW.  By an appropriate choice of the units of energy we can set $\kappa=2$.  
The magnitude of $u_s$ and $u_\rho$ can be adjusted to be anything we desire by appropriate rescaling of ${\bf S}_{\K}$ and $\rho_{\Q}$, respectively;  for historical reasons we take $u_s=u_\rho \equiv u =0.9$.
The sign of $\gamma_s$ and $\gamma_\rho$ is  significant;  positive $\gamma$ favors unidirectional (``stripe'') order while negative $\gamma$ favors bidirectional (``checkerboard'') order. We make the physically important assumption that unidirectional ordering is favored for both the SDW and CDW components.  However, the precise values of these couplings is not crucial, so we set $\gamma_s=\gamma_\rho \equiv \gamma$ and $\gamma=1$.  
$\tilde u_s$ and $\tilde \gamma_s$ are important only in determining whether the favored spin-density wave order is colinear (for $\tilde u_s$ large and positive) or spiral (for $\tilde u_s$ large and negative).  
In fact, colinear SDW order is favored even without the effect of these terms, so we take $\tilde u_s=\tilde\gamma_s=0$.  
We will always assume the interplane couplings are weak, and so only important in avoiding special features of the purely 2D limit; 
we further neglect the differences in the SDW and CDW interplane couplings and take  $J_{sz} = J_{\rho z} \equiv J_{z}$, and will 
take $J_z=0.0001$.
Finally $v$  represents the interaction between parallel components of the CDW and SDW orders, and $v^\prime$ the interaction between perpendicular components.  We assume these interactions are repulsive (positive) and that the interaction between parallel components is stronger than between perpendicular, $v> v^\prime$;  beyond that the specific values of these parameters are not extremely important so we take $v =1.5$ and $v^\prime=0.8$. 


\subsection{Mean-Field Transition Temperatures}

\begin{figure}[h]
    \begin{center}
    \includegraphics[width=3in]{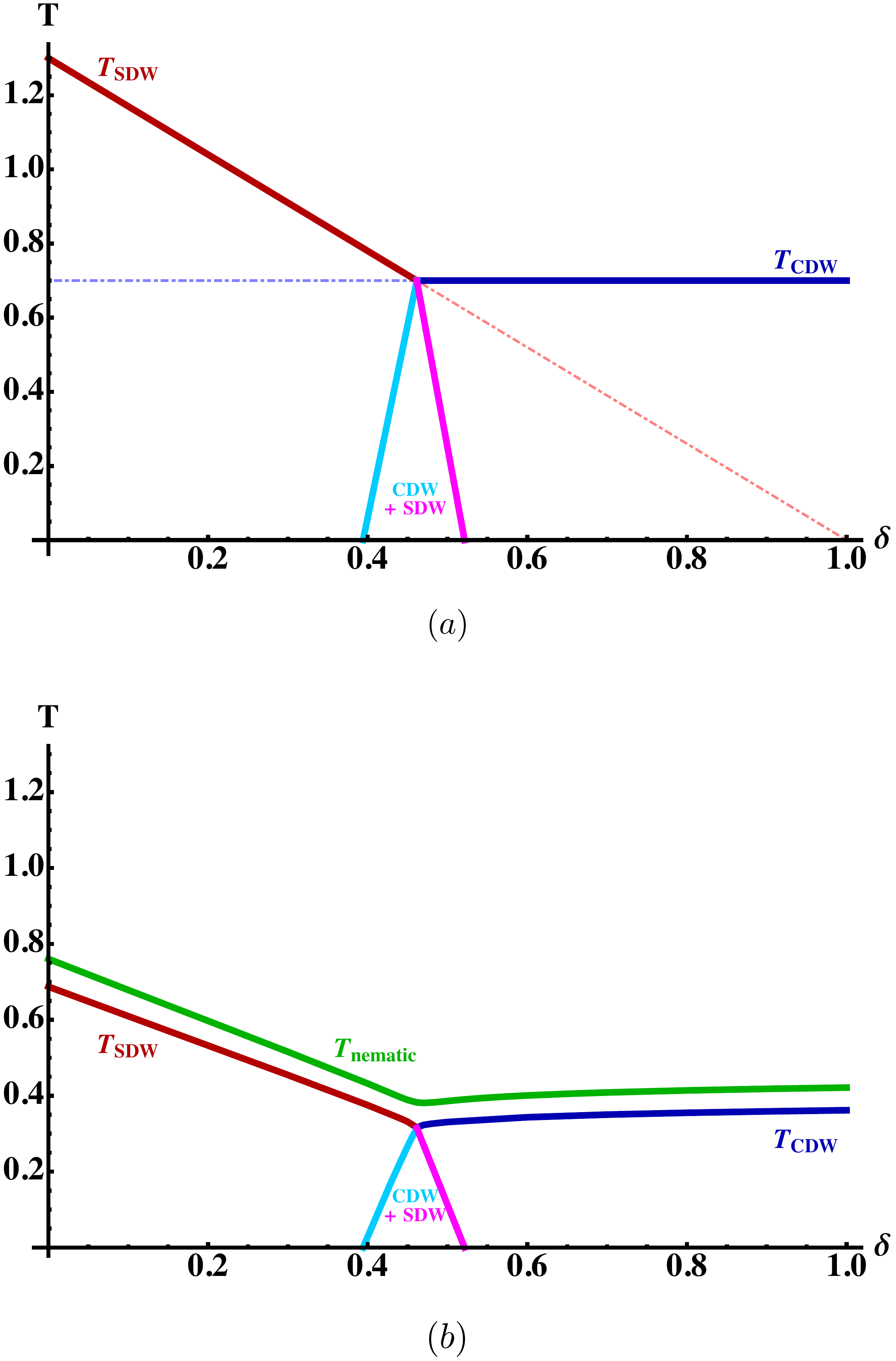}
    \end{center}

\caption{ (a) Mean-field phase diagram (ignoring spatial fluctuations) for the parameters defined in Sec. \ref{parameters}  and  Eqs.~\eqref{alphas} - ~\eqref{Tcdw} taken to be $\alpha_{s0} = \alpha_{\rho 0}=1.1$, $T_{\text{sdw}0} = 1.3$, $T_{\text{cdw}0} = 0.7$. Solid lines mark the second-order mean-field phase transitions. Red (blue) dot-dashed line marks the SDW (CDW) transition when there are no interaction terms between SDW and CDW, i.e., $v = v' = 0$.
(b) Large-$N$ phase diagram with the same parameters as in (a). The solid lines denote second order phase transitions and the four phase boundaries meet at a tetracritical point. If we had taken parameters such that $v\geq v' > \sqrt{u_s u_\rho}$, the narrow range of SDW and CDW coexistence at low temperature would have been replaced by a first order transition terminating at a bicritical point. The transition line to the (extremely narrow) vestigial nematic phase has been shifted upward by $\Delta T = 0.05$  for graphical clarity. }
\label{fig:MFandzerodisorder}
\end{figure}

The two remaining parameters in the effective field theory,
$\alpha_s$ and $\alpha_\rho$, tune the mean-field transitions of the density waves.  As is conventional, we assume them to be linearly varying functions of temperature, 
\be
\alpha_s = \alpha_{s0} \Big[T-T_{\text{sdw}}(\delta)  \Big], 
\label{alphas}
\ee
\be
\alpha_\rho = \alpha_{\rho 0}\Big[T-T_{\text{cdw}}(\delta) \Big], 
\label{alpharho}
\ee
where $T_{\text{sdw}}(\delta)$ and $T_{\text{cdw}}(\delta)$ are mean-field transition temperatures that are assumed to be functions of the doped hole concentration, (or some other control parameter) $\delta$. So as to illustrate the behavior in the neighborhood of a mean-field multicritical point (where $T_{\text{sdw}}(\delta)=T_{\text{cdw}}(\delta)$), we adopt the algebraically simple functions (shown in Fig.~\ref{fig:MFandzerodisorder}a)
\be
T_{\text{sdw}}(\delta) = T_{\text{sdw0}}(1-\delta),  
\label{Tsdw}
\ee
\be
T_{\text{cdw}}(\delta) = T_{\text{cdw0}} = const.
\label{Tcdw}
\ee
Note that because the Boltzmann weight is the exponential of $H/T$, the thermodynamic properties of the system have an explicit temperature dependence, as well as the implicit dependences implied by the above.  

\subsection{Coupling to disorder}
The most significant effect of disorder is to add a ``random field'' coupled to the CDW order parameters: 
\be
{\cal H}_{dis}=
h
\rho^*_{\Q} +  
h^\prime
 \rho^*_{\Q^\prime} + c.c.
\label{randomfield}
\ee
where $h(\vec r)$ is a Gaussian random variable which we take to be short-range correlated:
\bea
&&\overline{h(\vec r)}=0, \ \ \ \overline{h^*(\vec r)h(\vec r^\prime)}= \sigma^2 \delta_{n,n^\prime}\delta(x-x^\prime)\delta(y-y^\prime), \nonumber \\
&&\overline{h(\vec r)h(\vec r^\prime)}=\overline{h^*(\vec r)h^\prime(\vec r^\prime)}= \overline{h(\vec r)h^\prime(\vec r^\prime)}=0
\eea
and similarly for $h^\prime \leftrightarrow h$.
Time reversal symmetry precludes any similar term coupled to the SDW order parameter. Other forms of disorder coupling to the SDW order are permitted, including random mass disorder and random frustration (that can lead to a spin-glass phase), but, when weak, these are generally less important than random-field disorder.  We will return to this issue in the Sec. \ref{commensurability} below.


\subsection{Relative commensurability}

There is a special cubic term that couples the SDW and CDW order parameters: 
\bea
\label{cubic}
{\cal H}_{com} =&&\lambda \Big[\rho_{\Q}{\bf S}_{\K}^*\cdot {\bf S}_{\K}^* e^{i(\Q-2\K)\cdot\r}  \\
&&+\rho_{\Q^\prime}{\bf S}_{\K^\prime}^*\cdot {\bf S}_{\K^\prime}^* e^{i(\Q^\prime-2\K^\prime)\cdot\r} + {\rm c.c.}\Big] . \nonumber
\eea
Manifestly, this term is rapidly oscillating and so can be neglected unless either $2\K-\Q \equiv \vec 0$ ({\it i.e.} if the CDW and SDW are locked to be relatively commensurate) or in the presence of substantial disorder, where translation symmetry is no longer important. We will consider the significance of this term in Sec.~\ref{commensurability}, but for the purpose of the present analysis we assume that this term is negligible. 


\section{Variational and large $N$ solutions}
\label{variational}
To obtain an approximate solution for the phase digram of this effective field theory including the effects of fluctuations and, in particular, the possibility of partially melted phases (phases with vestigial order) we invoke the Feynman variational principal:
\be
F' \equiv F_{\text{trial}} + 
 \langle  H -  H_{\text{trial}} \rangle_{\text{trial}}  \ge F
\label{FreeEnergy}
\ee
where $F$ ($F_{\text{trial}}$) is the free energy corresponding to $H$  ($H_{\text{trial}}$), $\langle \dots  \rangle _{\text{trial}}$ denotes thermal average with Boltzmann weight $e^{-\beta H_{\text{trial}}}$. $H_{\text{trial}}$ is taken to be quadratic in fields with coefficients that are treated as variational parameters. Saddle-point equations are obtained by minimizing $F'$ with respect to the variational parameters. Details of the calculation can be found in Appendix~\ref{app:saddlepointequationsZerodisorder}.
 
The same saddle point equations can be viewed as the exact solution of a generalized version of the same problem in an appropriate large $N$ limit.  Here, we replace $\rho_{\Q}$ and $S_{\K}$ by corresponding $O(N)$ vectors, rescale the quartic terms by a factor of $1/N$, and solve the problem in the $N\to \infty$ limit.  An advantage of this latter approach is that systematic corrections to the saddle-point solution can, in principle, be computed in the context of a $1/N$ expansion.
 
 When considering the problem in the presence of disorder, we use the replica trick.  Specifically, we introduce $n$ replicas of the fluctuating fields, and then integrate out the quenched random variables, $h$ and $h^\prime$, as if they were thermal variables.  We solve the resulting replicated effective field theory using the same variational method as in the zero disorder case.  Finally, we take the replica limit, $n\to 0$. Details can  again be found in Appendix~\ref{app:saddlepointequationsFinitedisorder}.


\section{Results}
\label{results}
In this section we present three phase diagrams with a fixed set of input parameters but different disorder strengths: $\sigma=0$, 0.1, and 0.5. 


\subsection{Zero disorder}
$T_{\text{SDW}}$ and $T_{\text{CDW}}$, the transition temperatures for SDW and CDW respectively, are both suppressed when the effects of fluctuations are included, as shown in Fig.~\ref{fig:MFandzerodisorder}b. For $v > v'$, if $ v' <   \sqrt{u_s u_{\rho}}$ a coexisting phase occurs at low temperature with mutually perpendicular unidirectional SDW and CDW. This phase vanishes if $v > v' >  \sqrt{u_s u_{\rho}}$, when a first-order transition between SDW and CDW emerges, and a single nematic phase spanning both regimes.

For all values of $\delta$, the first ordered phase encountered on cooling from high temperatures is a nematic phase.  In this phase, both the CDW and the SDW orders vanish, 
$\langle \rho_{\Q}\rangle =0$ and $ \langle {\bf S}_{\K}\rangle ={\bf 0}$, but the fluctuations spectrum spontaneously breaks the $C_4$ rotational symmetry of the model.
In the variational treatment, the  nematic order parameters associated with vestigial SDW and CDW order are 
\be
\mathcal{N}_S = \langle |{\bf S}_{\K}|^2 - |{\bf S}_{\K^\prime}|^2  \rangle_{\text{trial}}
\label{NS}
\ee
and
\be
\mathcal{N}_\rho = \langle |\rho_{\Q}|^2 - |\rho_{\Q^\prime}|^2  \rangle_{\text{trial}},
\label{Nrho}
\ee
both of which are zero (by symmetry) at elevated temperatures and develop non-zero values at the nematic transition. Moreover, so long as other parameters are held fixed and the temperature is kept below $T_{\text{nematic}}$ as  $\delta$ is varied, $\mathcal{N}_S$ and $\mathcal{N}_N$ do not change in sign. 
In our case with $v>v'$, $\mathcal{N}_S$ and $\mathcal{N}_N$ always have opposite signs, $\mathcal{N}_S\mathcal{N}_N \leq 0$.  


\subsection{Effects of disorder}

\begin{figure}[h] 
    \begin{center}
    \includegraphics[width=3in]{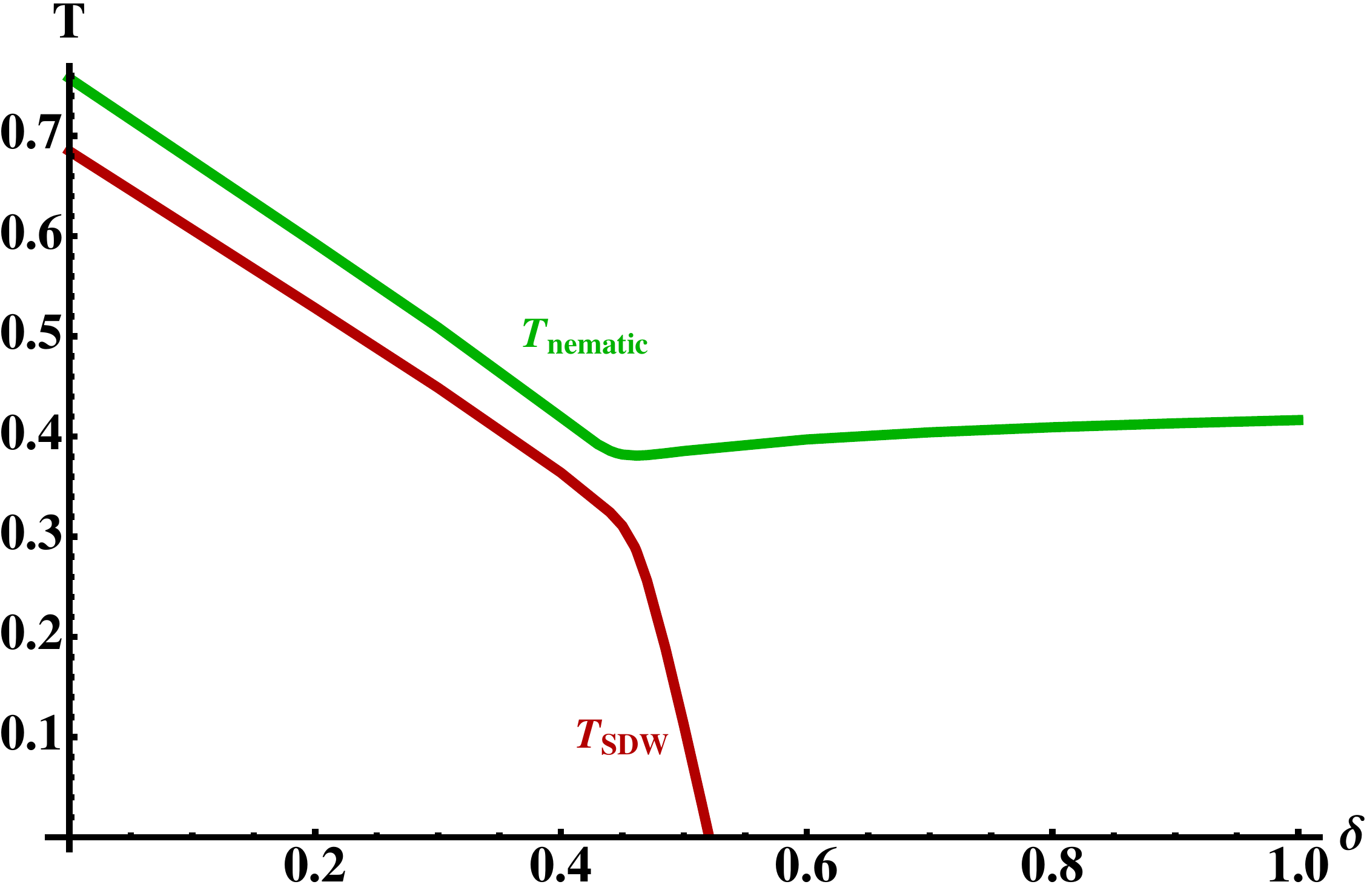}
    \end{center}
\caption{Large-$N$ phase diagram with input parameters same as in Fig.~\ref{fig:MFandzerodisorder}a, but with ``weak'' disorder, $\sigma=0.1$. All transitions are second-order. The nematic transition line is again shifted upward by $\Delta T = 0.05$ for graphical clarity. Consistent with general theorems, the CDW phase has been eliminated, but the nematic and SDW transition temperatures have only been decreased to almost unnoticeable degree relative to	the zero disorder case in Fig.~\ref{fig:MFandzerodisorder}b.}
\label{fig:phasediagramsigma01}
\end{figure}

The CDW transition will be suppressed when coupled to quenched random-field disorder.\cite{Imry-Ma-1975}
When disorder strength is weak, $\sigma=0.1$, the topology of the phase diagram is the similar to Fig.~\ref{fig:MFandzerodisorder}b except that there is no CDW phase, as shown in Fig.~\ref{fig:phasediagramsigma01}.  Neither the nematic nor the SDW order is much affected by this small amount of disorder.  This behavior can be understood on the basis of general theorems of statistical mechanics:  The lower critical dimension for the random field problem is $d_c=4$ for a continuous symmetry (e.g. the translation symmetry breaking associated with an incommensurate CDW) but $d_c=2$ for a discrete symmetry (e.g. the Ising nematic symmetry).  That the SDW order survives weak disorder is due to its non-trivial transformation under time reversal, which prevents the disorder potential from coupling like a conjugate field.  

\begin{figure}[h]
    \begin{center}
    \includegraphics[width=3in]{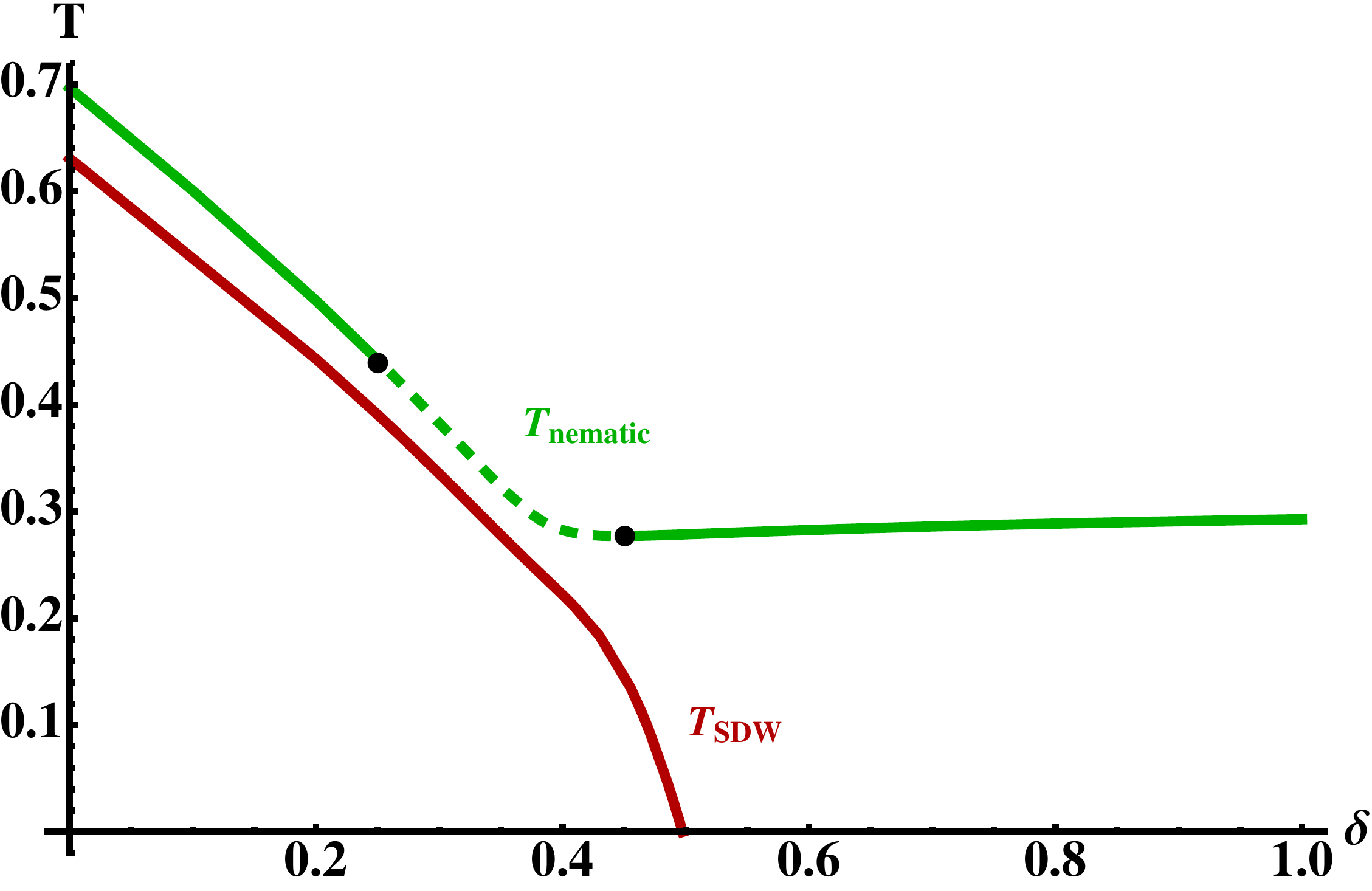}
    \end{center}
\caption{Large-$N$ phase diagram with the same parameters as in Fig.~\ref{fig:MFandzerodisorder}a, but with ``moderate'' disorder, $\sigma=0.5$. Dashed and solid green lines denote first- and second-order transitions respectively, and the black dots are tricritical points (see text). The nematic transition line is again shifted upward by $\Delta T = 0.05$ for graphical clarity. Notice the observable suppression of the nematic transition temperature compared to Fig.~\ref{fig:MFandzerodisorder}b and Fig.~\ref{fig:phasediagramsigma01}.
}
\label{fig:phasediagramsigma05}
\end{figure}

For slightly larger disorder, $\sigma=0.5$, the structure of the phase diagram changes somewhat as shown in Fig.~\ref{fig:phasediagramsigma05}.  Now, for a range of doping near where the multicritical point occured in the absence of disorder, the nematic transition has an altered character - it is first order due to the close proximity to the SDW phase,  
 with the two tricritical points at $\delta \approx 0.25$ and $\delta \approx 0.45$. 
 Within this doping range, we have confirmed at four selected values of $\delta$, both the first-order nature of the nematic transition and its separation from the second-order SDW transition. It is possible yet unlikely that a direct first-order transition occurs from isotropic phase to SDW phase at certain dopings within this regime.

We have not exhibited the behavior of the model at still larger disorder.  In a previous study~\cite{nie2014quenched} (which did not include an SDW order) we found that there was a substantially larger critical value of the disorder, $\sigma_c \approx 1.14$ (rescaled to be consistent with the current input parameters), beyond which the nematic phase no longer occurs, even in the limit $T\to 0$.  The presence of SDW fluctuations in the present model has a small quantitative effect on the magnitude of the critical disorder, but it remains the case that nematic order is quenched entirely for large enough $\sigma$.  We will not further discuss the effect of strong disorder on the SDW order, since in this context,  there is a more important effect of disorder  that arises indirectly from the here-to-for neglected coupling to the CDW order from Eq. \eqref{cubic}.

\section{Commensurate locking of the SDW and the CDW}
\label{commensurability}
The formal structure needed to treat the cubic term coupling the SDW and CDW fields, Eq. \eqref{cubic}, is beyond the scope of the straightforward variational approach taken in this paper.  Moreover, so long as $2\K-\Q$ is far from $\vec 0$, it's effects are negligible for most purposes.  We therefore defer the analysis of its effects to a future study.  However, there are two effects of this term which are necessary for the following discussion.

For mutually commensurate CDW and SDW order, this term has a determinative effect on the nature of the phase diagram, as can be seen already from a straightforward mean field theory, as in Ref. \onlinecite{zachar1998landau}.  If the preferred ordering vectors are close to commensurate, $|2\K-\Q| \ll 1$, then the interactions between the two orders is similar to the physics of the much studied commensurate to incommensurate (or Pokrovsky-Talapov) transition.\cite{pokrovsky-talapov-1979,pokrovsky-talapov-1980}  Typically what one expects in this case is a low temperature commensurately locked phase, with slightly shifted ordering vectors, ${\mathcal \K}$ and ${\mathcal \Q}$ with $2{\mathcal \K}\equiv {\mathcal \Q}$, but that above this temperature, the SDW and CDW ordering vectors relax to their preferred, not quite commensurate values.  

In the absence of disorder, if $|2\K-\Q|$ is sufficiently large, then this term is entirely negligible.  However, in the presence of disorder, the rapid phase oscillations of this term when integrated over space will not entirely eliminate its effectiveness.   To see this, imagine that to some degree $\rho_{\Q}$ is pinned by disorder, so there is an effective random field, $b^{\text{eff}}$:
\be 
b^{\text{eff}}(\vec r)=\lambda\ \rho_{\Q}(\vec r)\ e^{i(\Q-2\K)\cdot \vec r}
\label{beff}
\ee
that is conjugate to ${\bf S}_{\K}\cdot {\bf S}_{\K}$.  Although a weak random field of this sort  would not, generically, prevent the existence of a state that breaks spin-rotational symmetry (such as a spin-glass or a spin-nematic phase), because it couples to the phase of the SDW order, it is sufficient to preclude an ordered SDW phase in $D\leq 4$.  Presumably, this is responsible for the fact that the observed SDW phases in the cuprates always have a finite correlation length. In comparing the present theoretical results  (based on the effective field theory that neglects the effect of this coupling) with experiment, we will generally interpret experimentally observed phases with long magnetic correlation lengths as corresponding the ordered SDW states of the theory.


\section{Implications for the Cuprates}
\label{discussion}

We turn now to a discussion of the relevance of the present results to interpreting experiments in the cuprates. Here, we will focus primarily on YBCO and LSCO which we use as emblematic examples of the 123 and 214 families of hole doped cuprates.  We have not engaged in any fine-tuning of parameters - the behaviors we have documented are qualitatively robust as long as certain inequalities are satisfied.  


\subsection{YBCO}
Firstly, we summarize a set of observed properties  of the  much studied hole-doped cuprate YBCO which appear similar to the behaviors of the effective field theory we have presented above:

1) Although YBCO is orthorhombic, there is a clear $\delta$ dependent temperature below which various measures of crystalline anisotropy show a sudden nearly singular enhancement.\cite{ando-2002, achkar2016nematicity, lawler2010intra, fujita2014direct, hinkov2008electronic, daou-2010, louis2nematics}  We identify this as the nematic transition, $T_{\text{nematic}}$, rounded (and possibly pushed to slightly higher $T$) by the presence of a weak symmetry breaking field.

2) For $\delta$ between roughly 8\% and 15\%, there is clearly detectable short-range correlated CDW order below an onset temperature that is comparable to $T_{\text{nematic}}$.\cite{louis2nematics}  We identify this as the high doping regime of our phase diagram, where the nematic order is a vestige of the CDW. 

3) By contrast, for $\delta$ between  roughly 5\% and 8\%, CDW correlations have not been clearly identified, but strong SDW correlations - leading to quasi-static (``stripe-glass'') order at low $T$ - are clearly detectable.\cite{hinkov2008electronic, keimer2010NJP}  Now, the onset of SDW correlations appears to correlate with $T_{\text{nematic}}$, an observation that led to the suggestion that there are two thermodynamically distinct nematic phases.\cite{louis2nematics}  An important consequence of our  results is the implication that this is  a crossover between an SDW and a CDW dominated nematic.

4)  While the observed CDW correlations in the absence of a magnetic field are generally bidirectional, at high magnetic fields, where the CDW ordering tendency is enhanced (due to the suppression of superconductivity), the field induced CDW order is strongly unidirectional.  This suggests that the intrinsic ordering tendency is unidirectional, and that the bidirectional character seen at low fields is due to the effects of disorder.\cite{robertson2006distinguishing,nie2014quenched}  Note that the high field CDW order has its ordering vector, $\Q \sim 2\pi(0.3,0)$, oriented along the orthorhombic axis defined by the chain-direction in the crystal structure.  This is consistent with our results\cite{jang2016ideal} for $\gamma_\rho > 0$.

5)  In the   8\% to 15\% doping range, the SDW correlations are extremely short-ranged, and there is a substantial spin-gap.\cite{KeimerSpinGap-2000, DaiYBCOspinGap-2001}  The introduction of a small concentration of Zn impurities apparently pins  the  SDW fluctuations to some extent, and causes a corresponding suppression of the CDW order.\cite{keimerZn}  This confirms the fact that the SDW and CDW orders compete ($v$ and $v^\prime > 0$).  Moreover, the induced SDW correlations exhibit a preferred $\K$ that is far from commensurate with $\Q$, thus self-consistently justifying the neglect of the cubic coupling in Eq. \eqref{cubic}.

6)  In contrast, in the doping range 5\% and 8\%,  the unidirectional character of the SDW  below $T_{\text{nematic}}$ is observable\cite{hinkov2008electronic} even without the application of high fields.  It is notable that the preferred ordering vector, $\K \sim 2\pi(1,1-\delta)$, is not only mutually incommensurate with any observed CDW ordering vector, the incommensurability is in fact orthogonal to that of the CDW.  This latter observation is consistent with the natural hierarchy of couplings, $v> v^\prime>0$.


\subsection{The differences between YBCO and LSCO}
The major differences between the observed density wave phenomena in YBCO and LSCO can be attributed to the relevance of the cubic coupling between the CDW and SDW shown in Eq. \eqref{cubic}.  For given degree of mutual incommensurability, commensurate locking occurs only if the coupling constant  $\lambda$ exceeds a critical value $\lambda_c$, itself an increasing function of $|2\K-\Q|$.  Either because the preferred ordering vectors, $\K$ and $\Q$, are closer to being mutually commensurate in LSCO than in YBCO or because the magnitude of $\lambda$ is somewhat larger, the observed ordering vectors in LSCO always satisfy the commensurability relation ${\mathcal \Q}- 2{\mathcal \K}\equiv \vec 0$.  Because the term locking the CDW and SDW is cubic, when it is operational it supersedes the biquadratic couplings between the two orders that appear to dominate the physics in YBCO.  

There are two dramatic differences that result from this.  Firstly, it accounts for the fact that the interaction between the CDW and SDW orders in LSCO appears to be cooperative, while the two orders clearly compete in YBCO.  Secondly, it implies that the observed ordering vectors ${\mathcal \Q}$ and ${\mathcal \K}$ generically differ somewhat from the individually preferred ordering vectors, $\Q$ and $\K$.  However, by studying the fluctuating CDW and SDW order at elevated temperatures where they are no longer locked to one another, one can infer more accurately the preferred ordering vectors of each density wave separately.  

The validity of this perspective has been spectacularly supported by recent RIXS experiments of Dean and collaborators \cite{dean2017} on 1/8 doped LBCO (a stripe ordered member of the 214 family).
As is well known for this material,\cite{kivelson-rmp2003, FujitaStripe-2004, tranquada-2007} below the static spin-ordering transition temperatures, $T_{\text{SDW}}=42$K, the charge ordering vector is ${\mathcal \Q} =(\pm q,0)$ with $q=0.24$ and the spin ordering vector ${\mathcal \K}=(1/2\pm k,0)$ satisfies the mutual commensurability condition $2{\mathcal \K}\equiv {\mathcal \Q}$, {\it i.e.} $k=0.12$.\cite{commensurate}
Above $T_{\text{SDW}}$, no static spin order remains, but both fluctuating CDW and SDW correlations are still clearly distinguishable, albeit with correlation lengths and intensities that decrease with increasing $T$.  Most significant for present purposes is that the observed ordering vectors are temperature dependent in this regime:  The charge and spin ordering vectors cease to be mutually commensurate, $2{\mathcal \K} \neq {\mathcal \Q}$.  Specifically, $k$ decreases rapidly above $T_{\text{CDW}}$ to a ``high'' temperature value $k\approx 0.1$, while $q$ increases smoothly to a value $q\approx 0.272$.  

In 1/8 doped YBCO, $q\approx 0.32$ and $k\approx 0.1$ when this same material is lightly Zn doped (to pin the SDW fluctuations).\cite{keimerZn}  As Dean and collaborators have noted,\cite{dean2017} the greater similarity across families of hole-doped cuprates in the ordering tendencies at elevated temperatures strongly supports the notion that these tendencies are a robust, intrinsic feature of the electronic structure of the copper-oxide planes. Conversely, it is clear from our analysis that the relatively more dramatic family specific differences in the way these orders manifest at low temperatures can all be attributed to relatively small differences in the strength of a single term in the effective field theory.


\section{Additional Remarks}
\label{remarks}
We end with a few remarks about the broader context and implications of our results:

1) The strong asymmetry between the CDW and SDW portions of our calculated phase diagrams derives from the different way they couple to disorder.  Because the SDW is odd under time-reversal, no random-field-type coupling is allowed, and this would be true even were we to include the effects of spin-orbit coupling.  However, the SDW order is, of course, not impervious to disorder.  There is random $T_c$ disorder of the form $\delta\alpha_s(\vec r) |{\bf S}_{\K}|^2/2$, which in the present context is generated implicitly via the biquadratic coupling to the CDW order - $v$ and $v^\prime$.  However, if weak, such terms typically do not affect the structure of the phase diagram significantly,\cite{harris-lubensky-1974, cardy1996scaling} and indeed, since the Harris criterion\cite{harris1974effect} is satisfied, do not even affect the critical exponents (although they can lead to new multifractal correlations associated with rare events). As discussed in Sec. \ref{commensurability}, there  is also a higher order random field coupling, $b(\vec r){\bf S}_{\K}\cdot{\bf S}_{\K}+$ H.C., that is allowed by symmetry; this term is sensitive to the phase of the SDW order, and so even when weak, it eliminates  long-range SDW order in $D\leq 4$. We will discuss this in more detail in a forthcoming paper.

2) That there is a single nematic phase follows from symmetry.  Were we to integrate out the primary fields to derive an effective field theory in terms of nematic fields, ${\cal N}_S$ and ${\cal N}_\rho$ from Eqs. \eqref{NS} and \eqref{Nrho}, we would find that there is a symmetry allowed bilinear coupling, $\alpha_{S-\rho}\ \mathcal{N}_S \mathcal{N}_\rho$.  Consequently, when one form of nematic order occurs, the other necessarily occurs as well. 
However, other forms of the phase diagram are allowed:  Given that the microscopic character of the spin-driven nematic and the charge-driven nematic are so different, it would not be implausible that under some circumstances they could be separated by a first order transition, terminating in a critical or a bicritical point (although we have not found this in the range of parameters explored).

3)  Needless to say, the present discussion is based on rather general considerations, and is readily generalized to other circumstances.  For instance, in the Fe-based superconductors, there is a clearly observed Ising-nematic phase.  However, there is some uncertainty about whether this should be identified as being a form of  vestigial SDW order\cite{chen-yao-tsai-kivelson-2008,xu-muller-sachdev-2008} or associated with orbital ordering.\cite{lv-wu-phillips-2009,kruger-kumar-zaanen-van_den_brink-2009}  This discussion involves important microscopic considerations, but as the present analysis shows, it is likely that there is no sharp demarkation between the two behaviors.  Rather, we would expect as parameters are changed, there would occur a smooth crossover between an SDW dominated, to an orbital ordering dominated nematic phase without any intervening phase transitions. 

\begin{acknowledgements}
We thank Mark Dean, Brad Ramshaw, John Tranquada, Louis Taillefer, Gilles Tarjus, Yuxuan Wang, and Jian Kang for many useful discussions. This work was supported in part by the National Science Foundation grants DMR 1408713 at the University of Illinois (EF) and by the Department of Energy (DOE), Office of Science, Basic Energy Sciences, Materials Sciences and Engineering Division, under contract DE-AC02-76SF00515 at Stanford (LN,AM,SAK).
\end{acknowledgements}


\begin{widetext}
\appendix
\onecolumngrid

\section{Zero disorder}
\label{app:saddlepointequationsZerodisorder}
\renewcommand{\theequation}{A.\arabic{equation}}

We present the details of the variational approach and saddle-point equations.
The trial Hamiltonian $H_{\text{trial}}$ is taken to be
\bea
H_{\text{trial}} &\equiv &\int d\r  \Bigg\{ 
\frac{\kappa_{s\|}}{2} \Big[|\partial_x{\bf S}_{\K}|^2  +|\partial_y{\bf S}_{\K^\prime}|^2 \Big] + 
 \frac{\kappa_{s\perp}}{2} \Big[|\partial_y{\bf S}_{\K}|^2   +|\partial_x{\bf S}_{\K^\prime}|^2 \Big] + \frac {\tilde \alpha_{s}}2 |{\bf S}_{\K}|^2 +\frac {\tilde \alpha_{s}^\prime} 2 |{\bf S}_{\K^\prime}|^2 
     \nonumber\\
&& +  \frac{\kappa_{\rho\|}}{2} \Big[|\partial_x\rho_{\Q}|^2  +|\partial_y\rho_{\Q^\prime}|^2 \Big]+ 
 \frac{\kappa_{\rho\perp}}{2} \Big[|\partial_y\rho_{\Q}|^2   +|\partial_x\rho_{\Q^\prime}|^2 \Big] +\frac {\tilde \alpha_{\rho}}2 |\rho_{\Q}|^2 +\frac {\tilde \alpha_{\rho}^\prime} 2 |\rho_{\Q^\prime}|^2
 \nonumber\\
 && -J_{sz} \Big[ {\bf S}_{\K}(n) \cdot  {\bf S}^*_{\K}(n+1) + {\bf S}_{\K^\prime}(n) \cdot  {\bf S}^*_{\K^\prime}(n+1) +c.c. \Big] - J_{\rho z} \Big[  \rho_{\Q}(n)  \rho^*_{\Q}(n+1) + \rho_{\Q^\prime}(n)  \rho^*_{\Q^\prime}(n+1)+ c.c.   \Big]   \Bigg\}\nonumber\\
 && 
 \label{Htrial}
\eea
with variational parameters $\tilde \alpha_s,\tilde \alpha'_s, \tilde \alpha_{\rho}, \tilde \alpha'_{\rho}$. Minimization of $F'$ (Eq.~\eqref{FreeEnergy}) yields the saddle point equations:
\be
\tilde \alpha_s =  \alpha_s + \frac{u_s}{N} \Big(1+\frac{2}{N}\Big)  \langle |{\bf S}_{\K}|^2   \rangle +  \frac{(u_s+\gamma_s)}{N}\langle |{\bf S}_{\K'}|^2   \rangle  + \frac{v}{N}  \langle |\rho_{\Q}|^2  \rangle   + \frac{v'}{N}  \langle |\rho_{\Q'}|^2  \rangle
\label{saddle1}
\ee

\be
\tilde \alpha'_s =  \alpha_s + \frac{u_s}{N} \Big(1+\frac{2}{N}\Big)  \langle |{\bf S}_{\K'}|^2   \rangle +  \frac{(u_s+\gamma_s)}{N}\langle |{\bf S}_{\K}|^2   \rangle  + \frac{v}{N}  \langle |\rho_{\Q'}|^2  \rangle  +\frac{v'}{N}  \langle |\rho_{\Q}|^2  \rangle,
\label{saddle2}
\ee

\be
\tilde \alpha_{\rho} =  \alpha_\rho + \frac{u_\rho}{N} \Big(1+\frac{2}{N}\Big)  \langle  |\rho_{\Q}|^2   \rangle +  \frac{(u_\rho+\gamma_\rho)}{N}\langle  |\rho_{\Q'}|^2   \rangle  + \frac{v}{N}  \langle |{\bf S}_{\K}|^2  \rangle   + \frac{v'}{N}  \langle |{\bf S}_{\K'}|^2   \rangle,
\label{saddle3}
\ee

\be
\tilde \alpha'_{\rho} =  \alpha_\rho + \frac{u_\rho}{N} \Big(1+\frac{2}{N}\Big)  \langle  |\rho_{\Q'}|^2   \rangle +  \frac{(u_\rho+\gamma_\rho)}{N}\langle  |\rho_{\Q}|^2   \rangle  + \frac{v}{N}  \langle |{\bf S}_{\K'}|^2  \rangle   + \frac{v'}{N}  \langle |{\bf S}_{\K}|^2   \rangle,
\label{saddle4}
\ee
where for clarity we restore the factor of $1/N$ before taking the limit of $N \to \infty$, 
and adopt the short-hand notation $\langle \dots \rangle \equiv \langle \dots \rangle_{\text{trial}}$, and 
\be
\langle |{\bf S}_{\K}|^2  \rangle = \frac{1}{2}  \int \frac{d \vec k}{(2\pi)^3} \frac{NT}{  \frac{\kappa_{s\|}}{2} k^2_x + \frac{\kappa_{s\perp}}{2} k^2_y  - 2J_{sz} \cos k_z  + \frac{\tilde \alpha_s}{2}   },
\label{Sq}
\ee

\be
\langle |{\bf S}_{\K'}|^2  \rangle = \frac{1}{2}   \int \frac{d \vec k}{(2\pi)^3} \frac{NT}{  \frac{\kappa_{s\|}}{2} k^2_y + \frac{\kappa_{s\perp}}{2} k^2_x  - 2J_{sz} \cos k_z  + \frac{\tilde \alpha'_s}{2}   },
\label{Sqprime}
\ee

\be
\langle |\rho_{\Q}|^2 \rangle = \frac{1}{2}    \int \frac{d \vec k}{(2\pi)^3} \frac{NT}{  \frac{\kappa_{\rho\|}}{2} k^2_x + \frac{\kappa_{\rho\perp}}{2} k^2_y  - 2J_{\rho z} \cos k_z  + \frac{\tilde \alpha_\rho}{2}   },
\label{rhok}
\ee

\be
\langle |\rho_{\Q'}|^2 \rangle =  \frac{1}{2}   \int \frac{d \vec k}{(2\pi)^3} \frac{NT}{  \frac{\kappa_{\rho\|}}{2} k^2_y + \frac{\kappa_{\rho\perp}}{2} k^2_x  - 2J_{\rho z} \cos k_z  + \frac{\tilde \alpha'_\rho}{2}   },
\label{rhokprime}
\ee
where $T=1/\beta$ is the temperature. The four variational parameters are solved self-consistently from Eqs.~\eqref{saddle1} to~\eqref{saddle4}. 


\section{Finite disorder}
\label{app:saddlepointequationsFinitedisorder}
\renewcommand{\theequation}{B.\arabic{equation}}

Here we present the derivation of saddle-point equations with finite disorder. Applying replica trick to the original Hamiltonian Eq.\eqref{Hamiltonian} and integrating out $h$ and $h^\prime$, we obtain the replicated Hamiltonian:
\bea
\tilde H &=&  \sum\limits_{a=1}^M \int d\r   \Bigg\{ 
\frac{\kappa_{s\|}}{2} \Big[|\partial_x{\bf S}_{\K,a}|^2  +|\partial_y{\bf S}_{\K^\prime,a}|^2 \Big] + 
 \frac{\kappa_{s\perp}}{2} \Big[|\partial_y{\bf S}_{\K,a}|^2   +|\partial_x{\bf S}_{\K^\prime,a}|^2 \Big]   +\frac {\alpha_s} 2 \Big[ |{\bf S}_{\K,a}|^2  +|{\bf S}_{\K^\prime,a}|^2 \Big]      \nonumber\\
     && \qquad  \qquad +  \frac{\kappa_{\rho\|}}{2} \Big[|\partial_x\rho_{\Q,a}|^2  +|\partial_y\rho_{\Q^\prime,a}|^2 \Big]+ 
 \frac{\kappa_{\rho\perp}}{2} \Big[|\partial_y\rho_{\Q,a}|^2   +|\partial_x\rho_{\Q^\prime,a}|^2 \Big] +\frac {\alpha_\rho} 2 \Big[ |\rho_{\Q,a}|^2 +|\rho_{\Q^\prime,a}|^2 \Big] \nonumber\\
 && \qquad  \qquad -J_{sz} \Big[ {\bf S}_{\K,a}(n) \cdot  {\bf S}^*_{\K,a}(n+1) + {\bf S}_{\K^\prime,a}(n) \cdot  {\bf S}^*_{\K^\prime,a}(n+1) +c.c. \Big]   \nonumber\\
&& \qquad  \qquad - J_{\rho z} \Big[  \rho_{\Q,a}(n)  \rho^*_{\Q,a}(n+1) + \rho_{\Q^\prime,a}(n)  \rho^*_{\Q^\prime,a}(n+1)+ c.c.   \Big] \nonumber\\
 &&  \qquad  \qquad  + \frac {u_s}{4N} \Big[ |{\bf S}_{\K,a}|^2   +|{\bf S}_{\K^\prime,a}|^2 \Big]^2 
+ \frac {\gamma_s}{2N}  |{\bf S}_{\K,a}|^2 |{\bf S}_{\K^\prime,a}|^2  
+ \frac {u_\rho}{4N} \Big[ |\rho_{\Q,a}|^2 +|\rho_{\Q^\prime,a}|^2 \Big]^2  + \frac {\gamma_\rho} {2N} |\rho_{\Q,a}|^2 |\rho_{\Q^\prime,a}|^2   \nonumber\\
 &&  \qquad  \qquad  + \frac{v}{ 2N} \Big[|\rho_{\Q,a}|^2|{\bf S}_{\K,a}|^2+|\rho_{\Q^\prime,a}|^2|{\bf S}_{\K^\prime,a}|^2\Big] + \frac {v'}{2N} \Big[|\rho_{\Q,a}|^2|{\bf S}_{\K',a}|^2+|\rho_{\Q^\prime,a}|^2|{\bf S}_{\K,a}|^2\Big]   \Bigg\}  \nonumber\\
  && \qquad  \qquad+  \sum\limits_{a,b=1}^M \int d\vec r  \ \ \Big(-\frac{2\sigma^2}{T} \Big)    \Big( \rho^*_{\Q,a} \rho_{\Q,b} + \rho^*_{\Q',a} \rho_{\Q',b} \Big),
 \label{Hreplica}
\eea
where the appropriate factors of $1/N$ were added, $a,b$ are replica indices, $M$ is the total number of replicas. Similar to the zero-disorder case, we introduce a quadratic trial Hamiltonian with four variational parameters $\tilde \alpha_s, \tilde \alpha_s', \tilde \alpha_\rho, \tilde\alpha_\rho'$:
\bea
\tilde H_{\text{trial}} &\equiv& \sum\limits_{a=1}^M \int d\r  \Bigg\{ 
\frac{\kappa_{s\|}}{2} \Big[|\partial_x{\bf S}_{\K,a}|^2  +|\partial_y{\bf S}_{\K^\prime,a}|^2 \Big] + 
 \frac{\kappa_{s\perp}}{2} \Big[|\partial_y{\bf S}_{\K,a}|^2   +|\partial_x{\bf S}_{\K^\prime,a}|^2 \Big] + \frac {\tilde \alpha_{s}}2 |{\bf S}_{\K,a}|^2 +\frac {\tilde \alpha_{s}^\prime} 2 |{\bf S}_{\K^\prime,a}|^2 
     \nonumber\\
&& \qquad  \qquad +  \frac{\kappa_{\rho\|}}{2} \Big[|\partial_x\rho_{\Q,a}|^2  +|\partial_y\rho_{\Q^\prime,a}|^2 \Big]+ 
 \frac{\kappa_{\rho\perp}}{2} \Big[|\partial_y\rho_{\Q,a}|^2   +|\partial_x\rho_{\Q^\prime,a}|^2 \Big] +\frac {\tilde \alpha_{\rho}}2 |\rho_{\Q,a}|^2 +\frac {\tilde \alpha_{\rho}^\prime} 2 |\rho_{\Q^\prime,a}|^2  \nonumber\\
 && \qquad \qquad -J_{sz} \Big[ {\bf S}_{\K,a}(n) \cdot  {\bf S}^*_{\K,a}(n+1) + {\bf S}_{\K^\prime,a}(n) \cdot  {\bf S}^*_{\K^\prime,a}(n+1) +c.c. \Big] \nonumber\\
&& \qquad \qquad - J_{\rho z} \Big[  \rho_{\Q,a}(n)  \rho^*_{\Q,a}(n+1) + \rho_{\Q^\prime,a}(n)  \rho^*_{\Q^\prime,a}(n+1)+ c.c.   \Big]  \Bigg\} \nonumber\\
&&  \qquad  \quad  +\sum\limits_{a,b=1}^M   \int d\vec r \ \   \Big(-\frac{2\sigma^2}{T} \Big)  \Big( \rho^*_{\Q,a} \rho_{\Q,b} + \rho^*_{\Q',a} \rho_{\Q',b} \Big).
\label{Hreplicatrial}
\eea
The saddle-point equations remain formally the same as~\eqref{saddle1} $\sim$ ~\eqref{saddle4}, but with $\langle \dots  \rangle$ replaced by $\overline{\langle \dots  \rangle}$, where the overline denotes disorder configuration average. Specifically, 
\be
\overline{\langle |\rho_{\vec Q,a}|^2  \rangle} = \frac{1}{2}  \int \frac{d \vec k}{(2\pi)^3} \frac{NT}{  \frac{\kappa_{\rho\|}}{2} k^2_x + \frac{\kappa_{\rho\perp}}{2} k^2_y  - 2J_{\rho z} \cos k_z  + \frac{\tilde \alpha_\rho}{2}   }     +     \int \frac{d \vec k}{(2\pi)^3} \frac{N\sigma^2}{  \Big(  \frac{\kappa_{\rho \|}}{2} k^2_x + \frac{\kappa_{\rho \perp}}{2} k^2_y  - 2J_{\rho z} \cos k_z  + \frac{\tilde \alpha_{\rho}}{2}   \Big)^2},
  \label{rhokdisorder}
\ee
\be
\overline{\langle |\rho_{\vec Q,a}|^2  \rangle} = \frac{1}{2}  \int \frac{d \vec k}{(2\pi)^3} \frac{NT}{  \frac{\kappa_{\rho\|}}{2} k^2_y + \frac{\kappa_{\rho\perp}}{2} k^2_x  - 2J_{\rho z} \cos k_z  + \frac{\tilde \alpha_\rho'}{2}   }     +     \int \frac{d \vec k}{(2\pi)^3} \frac{N\sigma^2}{  \Big(  \frac{\kappa_{\rho \|}}{2} k^2_y + \frac{\kappa_{\rho \perp}}{2} k^2_x  - 2J_{\rho z} \cos k_z  + \frac{\tilde \alpha_{\rho'}}{2}   \Big)^2},
  \label{rhokprimedisorder}
\ee
meanwhile $\langle |{\bf S}_{\K}|^2  \rangle$ and $\langle |{\bf S}_{\K^\prime}|^2  \rangle$ remain the same as~\eqref{Sq} and~\eqref{Sqprime}.

\end{widetext}

\bibliographystyle{apsrev4-1} 
\bibliography{refs}

\begin{thebibliography}{69}
\bibitem [{su2()}]{su2}%
  \BibitemOpen
  \href@noop {} {}\bibinfo {note} {{In writing the effective field theory in
  Eq. \eqref{Hamiltonian}, we have assumed an underlying spin $SU(2)$ symmetry.
  If we include the effects of spin-orbit coupling, there are a number of
  additional terms which are symmetry allowed.}}\BibitemShut {Stop}%
\bibitem [{com()}]{commensurate}%
  \BibitemOpen
  \href@noop {} {}\bibinfo {note} {{Note that neither vector is strictly
  commensurate with the underlying lattice, {\it i.e.} the difference between
  0.23 and 1/4 is large compared to the experimental uncertainty in the
  ordering vector.}}\BibitemShut {Stop}%
\end{thebibliography}

\end{document}